\documentclass [reprint,aps,amsmath,amssymb,twoside,prl,showkeys,superscriptaddress]{revtex4-1}
\usepackage{verbatim}
\usepackage{multirow}
\usepackage{graphicx}
\usepackage[T1]{fontenc}
\usepackage{textcomp}
\usepackage{txfonts}
\usepackage{tensor}
\usepackage{cancel}
\usepackage{color}
\newcommand{\HCd}{\mathcal{H}}
\newcommand{\HCdtD}{\tilde{\HCd}_{\mathrm{Dyn}}}

\def\HCdtD{\tilde{\HCd}_{\mathrm{Dyn}}}

\def\HCdt0{\tilde{\HCd}_{0}}

\newcommand{\dd}{\,\mathrm{d}}

\newcommand{\onehalf}{{\textstyle\frac{1}{2}}}
\newcommand{\onethird}{{\textstyle\frac{1}{3}}}

\newcommand{\quarter}{{\textstyle\frac{1}{4}}}
\newcommand{\pfrac}[2]{\frac{\partial{#1}}{\partial{#2}}}

\newcommand{\ddfrac}[2]{\frac{\dd#1}{\dd#2}}

\newcommand{\afffias}{Frankfurt Institute for Advanced Studies (FIAS), Ruth-Moufang-Strasse~1, 60438 Frankfurt am Main, Germany}
\newcommand{\affjwg}{Goethe-Universit\"at, Max-von-Laue-Strasse~1, 60438~Frankfurt am Main, Germany}
\newcommand{\affgsi}{GSI Helmholtzzentrum f\"ur Schwerionenforschung GmbH, Planckstrasse~1, 64291 Darmstadt, Germany}
\newcommand{\affbgu}{Physics Department, Ben-Gurion University of the Negev, Beer-Sheva 84105, Israel}
\newcommand{\affbahamas}{Bahamas Advanced Study Institute and Conferences, 4A Ocean Heights, Hill View Circle, Stella Maris, Long Island, The Bahamas}
\begin{document}
\title{Energy transfer from space-time into matter and a bouncing inflation\\ from Covariant Canonical Gauge theory of Gravity}
\author{D.~Benisty}\email{benidav@post.bgu.ac.il}
\affiliation{\afffias}\affiliation{\affjwg}\affiliation{\affbgu}
\author{D.~Vasak}\email{vasak@fias.uni-frankfurt.de}
\affiliation{\afffias}
\author{E.I. ~Guendelman}
\affiliation{\afffias}\affiliation{\affbgu}\affiliation{\affbahamas}\email{guendelman@fias.uni-frankfurt.de}
\author{J.~Struckmeier}\email{struckmeier@fias.uni-frankfurt.de}
\affiliation{\afffias}\affiliation{\affjwg}\affiliation{\affgsi}
\received{\today}
%
\begin{abstract}
Cosmological solutions for covariant canonical gauge theories of gravity are presented. The underlying covariant canonical transformation framework invokes a dynamical space-time Hamiltonian consisting of the Einstein-Hilbert term plus a quadratic Riemann tensor invariant with a fundamental dimensionless coupling constant $g_1$. A typical time scale related to this constant, $\tau = \sqrt{8 \pi G g_1}$, is characteristic for the type of cosmological solutions: for $t \ll \tau$ the quadratic term is dominant, the energy momentum tensor of matter is not covariantly conserved, and we observe modified dynamics of matter and space-time. On the other hand, for $t \gg \tau$, the Einstein term  dominates and the solution converges to classical cosmology. This is analyzed for different types of matter and dark energy with a constant equation of state. While for a radiation dominated universe solution the cosmology does not change, we find for a dark energy universe the well known de-Sitter space. However, we also identify a special  bouncing solution (for $k=0$) which for large times approaches the de-Sitter space again. For a dust dominated universe  (with no pressure) deviations are seen only in the early epoch. In late epoch the solution asymptotically behaves as the standard dust solution. 
%
%
\end{abstract}
\maketitle
\section{Introduction}
%
%
The Covariant Canonical Gauge Gravitation (CCGG) is a theory derived from the canonical transformation theory in the Hamiltonian picture \cite{Struckmeier:2017vkf}\cite{Struckmeier:2017stl}. Starting of with matter fields embedded in a dynamical metric, it naturally identifies the affine connection as the fundamental gauge field. The theory includes torsion, as the affine connection does not have to be symmetric. In addition, it complements the Einstein Hilbert action by an additional quadratic Riemann invariant that formally corresponds to a squared momentum field equipping space-time with kinetic energy. In consequence the covariant conservation of the stress energy tensor is violated, as many modified theories of gravity in the Palatini approach do, where the connection is being an independent degree of freedom \cite{Palatini}. For a symmetric connection, we got a relation between the covariant conservation of the energy momentum tensor and the metricity condition \cite{Benisty:2018efx}\cite{Benisty:2018ufz}. With the torsion invoked, though, the covariant conservation of the energy momentum tensor is violated.

The objective of this paper is to investigate the impact of that quadratic term on the dynamics of the universe, and the behavior of matter, in the Friedman model. We use natural units with $\hbar = c = 1$. The signature of the metric is $g_{\mu\nu} = \mathrm{diag} (1,-1,-1,-1)$. Small Greek indices run from 0 to 3 and denote the number of space-time dimensions; Small Latin indices run from 1 to 3 and denote spatial dimensions only. 

Different Gauge theories of gravity based on the local Poincare group, instead of the general coordinate invariant group has been extensively studies in Ref \cite{Hehl:1994ue}-\cite{Chen:2009at}.

\section{Covariant Canonical Gauge Theory of Gravity}
A closed description of the coupled dynamics of fields and space-time geometry has been
derived in \cite{Struckmeier:2017vkf}. The CCGG formalism results in an amended  covariant Hamiltonian density $\mathcal{H}_m$ of the involved matter fields, to which a linear-quadratic Hamiltonian $\HCdtD$ of "free" space-time is added:
\begin{align}
\tilde{\mathcal{H}} &= \HCdtD + \tilde{\mathcal{H}}_m \\
&=\frac{1}{4g_{1}}\tilde{q}\indices{_{\eta}^{\alpha\xi\beta}}
q\indices{_{\alpha}^{\eta\tau\lambda}}\,g_{\xi\tau}g_{\beta\lambda}-
g_{2}\tilde{q}\indices{_{\eta}^{\alpha\eta\beta}}g_{\alpha\beta}+g_{3}\sqrt{-g} + \tilde{\mathcal{H}}_m. \nonumber
\label{action}
\end{align}
Tilde denotes multiplication by $\sqrt[]{-g}$. The tensor densities $\tilde{k}^{\,\alpha\lambda\beta}$ and $\tilde{q}\indices{_{\eta}^{\alpha\xi\beta}}$ are the canonical momenta conjugate to the metric field, $g_{\alpha\lambda}$, and to the connection coefficient field $\gamma\indices{^{\eta}_{\alpha\xi}}$,  respectively. The energy-momentum tensor of matter, the stress tensor, is defined as
\begin{equation}
\theta^{\mu\nu} := \frac{2}{\sqrt[]{-g}} \pfrac{\tilde{\mathcal{H}}_m}{g_{\mu\nu}}.
\end{equation}
In analogy, the energy-momentum (strain) tensor of space-time, 
\begin{equation}\label{eq:final-lag}
\Theta^{\mu\nu} := \frac{2}{\sqrt[]{-g}} \pfrac{\HCdtD}{g_{\mu\nu}} = \frac{1}{8\pi G}G\indices{^{\mu\nu}}-g_{1} Q\indices{^{\mu\nu}},
\end{equation}
is built from the Einstein tensor
\begin{equation}
\label{eq:final-lag1}G\indices{^{\alpha}_{\xi}}:= R\indices{^{\alpha}_{\xi}}-
\onehalf\delta_{\xi}^{\alpha}R
\end{equation}
and the contribution from the quadratic Riemann term denoted by
\begin{equation}\label{eq:r-squared}
Q\indices{^{\alpha}_{\xi}}:= R\indices{^{\eta\beta\lambda\alpha}}R\indices{_{\eta\beta\lambda\xi}}-
\quarter\delta_{\xi}^{\alpha}R\indices{^{\eta\beta\lambda\sigma}}
R\indices{_{\eta\beta\lambda\sigma}}.
\end{equation}
Hereby we deploy the standard notations  
$R := R^{\mu\nu} g_{\mu\nu}$ for the Ricci scalar, $R^{\mu\nu} := R\indices{_\alpha^{\mu\alpha\nu}}$ for the Ricci tensor, and $R^{\mu\nu\alpha\beta}$ for the Riemann tensor.

With these definitions, the CCGG equation, generalizing the Einstein equation, can be written as a balance equation of the stress and strain tensors, 
\begin{equation}
\Theta\indices{_\mu^\nu} + \theta\indices{_\mu^\nu} = 0. 
\label{eq:balance}
\end{equation}
Of course, by setting $g_1 = 0$ this equation reduces to the Einstein equation with  cosmological constant $\Lambda$. This is easy to see as then in Eq.~(\ref{eq:final-lag}) only the linear term remains, and the fundamental coupling constants of the CCGG Hamiltonian map to the gravitational coupling constant, $G$, and the cosmological constant $\Lambda$ according to 
\begin{equation}
g_1 g_2 = \frac{1}{16 \pi G} \quad , \quad 6g_1 g_2^2 + g_3 = \frac{\Lambda}{8 \pi G}.
\end{equation}
As discussed in Ref. \cite{Vasak:2018gqn}, this fundamental Hamiltonian density leads to a new interpretation of the  cosmological constant, and its richer structure compared to Einstein-Hilbert allows to resolve the so called Cosmological Constant Problem.

\subsection{The stress energy tensor conservation}
The covariant divergence of the stress energy-momentum 
tensor does not vanish for this formulation, and does not provide a covariantly conserved stress tensor:
\begin{equation}
\theta^{\mu\nu}_{\text{total};\nu} \neq 0.
\end{equation}
This can be seen as follows. While the covariant divergence of the Einstein tensor vanishes identically, $G\indices{^{\mu\nu}_{;\nu}}=0$, this is not the case for the quadratic term in the strain tensor. A direct calculation leads (see Appendix A) to
\begin{equation}
Q\indices{^{\alpha}_{\xi;\alpha}}=R\indices{^{\,\eta\beta\lambda\alpha}_{;\alpha}}\,R\indices{_{\eta\beta\lambda\xi}} \neq 0.
\end{equation}
Hence
\begin{equation} \label{eq:thetaderne0}
-\Theta\indices{^{\mu\nu}_{;\nu}} \equiv g_{1}\,Q\indices{^{\mu\nu}_{;\nu}} = \theta\indices{^{\mu\nu}_{;\nu}} \ne 0.
\end{equation}

This implies that we should expect energy and momentum transfer between space-time (described by the quadratic Riemann term in the strain tensor) and the energy-momentum density of matter, described by the stress tensor. We will come back to this conjecture below. In the following we illuminate using the Friedman model how the CCGG theory modifies cosmology. 
\section{Cosmological implications}
\subsection{Generalized Friedman equation}
The (FLRW) Friedman-Lemaitre-Robertson-Walker ansatz \cite{Friedman:1922kd} is the standard model of cosmology dynamics based on the assumption of a homogeneous and isotropic universe at any point, commonly referred to as the cosmological principle. The symmetry considerations lead to the FLRW metric
\begin{equation}\label{eq:robwalk}
\dd s^{2}=\dd t^{2}-a^{2}(t)\left[\frac{\dd r^{2}}{1-Kr^{2}}+r^{2}\left(\dd\theta^{2}+\sin^{2}\theta\dd\phi^{2}\right)\right].
\end{equation}
Herein, $a(t)$ defines the dimensionless cosmological expansion (scale) factor,
whereas $K$ denotes the positive, negative, or zero special curvature $K$ of the spatial slice.
In the following, we determine the expansion factor dynamics $a(t)$ by means of our generalized field equation (\ref{eq:balance}).
To set up the source term, the universe is usually modeled as a \emph{perfect fluid}.
The appropriate energy-momentum tensor is then
\begin{equation}\label{eq:robwalk-T}
\theta\indices{^{\alpha}_{\xi}}=\mathrm{diag}(\rho,-p,-p,-p).
\end{equation}
The density, $\rho$, and the pressure, $p$, refer to all types of matter present in the universe. Due to the symmetry properties they can only depend on the universal time $t$. 
The density $\rho(t)$ and the pressure $p(t)$ are not independent but related via an \emph{equation of state}, which, 
for a perfect fluid, is characterized by a constant parameter $\omega$:
\begin{equation}\label{eq:eos}
\omega = \frac{p}{\rho} \, \qquad \omega=\mathrm{const}.
\end{equation}
As the trace of the quadratic Riemann tensor vanishes, it does not contribute to the trace equation. So independently of the dimensionless constant $g_{1}$, which is associated with the quadratic Riemann tensor terms, the metric~(\ref{eq:robwalk}) yields the following inhomogeneous second-order equation for the expansion factor:
\begin{equation}\label{eq:robwalk-trace}
\frac{\ddot{a}}{a}+\left(\frac{\dot{a}}{a}\right)^2-2M+\frac{K}{a^2}=0,
\end{equation}
with
\begin{equation}\label{eq:M-squared}
M(t)=\textstyle\frac{1}{3}\left[2\pi G\,\left(\,\rho(t)-3p(t)\right)+\Lambda\right].
\end{equation}
On the basis of the metric~(\ref{eq:robwalk}), the non-contracted
equation~(\ref{eq:balance}) yields two more differential equation for
the expansion factor $a(t)$ for the indices $\alpha,\xi=0$ and $\alpha,\xi=1$
\begin{subequations}\label{eq:friedmann}
\begin{equation}
-8\pi Gg_{1}\left[\left(\frac{\dot{a}^{2}+K}{a}\right)^{2}-\ddot{a}^{2}\right]+
\dot{a}^{2}+K-\onethird\Lambda a^{2} =\frac{8\pi G}{3}\rho a^{2},
\end{equation}
\begin{equation}
8\pi Gg_{1}\left[\left(\frac{\dot{a}^{2}+K}{a}\right)^{2}-\ddot{a}^{2}\right]+
2a\ddot{a}+\dot{a}^{2}+K-\Lambda a^{2} =-8\pi Gpa^{2}.
\end{equation}
\end{subequations}
For those generalized Friedman equations, we first consider two asymptotic  cases. Firstly, for $g_{1}=0$,
\begin{subequations}\label{eq:friedmannor}
\begin{equation}
\left(\frac{\dot{a}}{a}\right)^{2}+\frac{K}{a^2}-\onethird\Lambda =\frac{8\pi G}{3}\rho,
\end{equation}
\begin{equation}
2\frac{\ddot{a}}{a}+\left(\frac{\dot{a}}{a}\right)^{2}+\frac{K}{a^2}-\Lambda =-8\pi G p,
\end{equation}
\end{subequations}
and we recover the conventional Friedman equations. In the second extreme case, $g_1 \rightarrow \infty$, the contribution of the $g_1$ terms dominates the Friedman equations, and 
\begin{subequations}\label{eq:friedmanninf}
\begin{equation}
-g_1\left[\left(\frac{\dot{a}^{2}+K}{a}\right)^{2}-\ddot{a}^{2}\right]=\frac{1}{3}\rho_{g_1\rightarrow \infty} a^{2},
\end{equation}
\begin{equation}
-g_{1}\left[\left(\frac{\dot{a}^{2}+K}{a}\right)^{2}-\ddot{a}^{2}\right] =p_{g_1\rightarrow \infty}a^{2}.
\end{equation}
\end{subequations} 
In this limit the equation of state, defined as
\begin{equation}\label{eofr}
\omega_{g_1\rightarrow \infty} \equiv \frac {p_{g_1\rightarrow \infty}}{\rho_{g_1\rightarrow \infty}},
\end{equation}
is that of radiation, $\omega_{g_1\rightarrow \infty} = \frac{1}{3}$. This is not surprising because now the strain tensor is dominated by the quadratic term which is traceless as in the case of radiation.

For arbitrary $g_1$, Eqs.~(\ref{eq:robwalk-trace}) and~(\ref{eq:friedmann}) must hold simultaneously. We can resolve Eq.~(\ref{eq:robwalk-trace}) for $\ddot{a}$ and insert it into Eq.~(\ref{eq:friedmann}):
\begin{equation}\label{eq:friedmann3}
32\pi G\,g_{1}M\left(\dot{a}^{2}+K-Ma^{2}\right)+
\dot{a}^{2}+K-\onethird\Lambda a^{2}=\frac{8\pi G}{3}\rho a^{2}.
\end{equation}
The term proportional to $g_{1}$ thus gives rise to modified cosmic dynamics,
as compared to the conventional Friedman equation, to which
Eq.~(\ref{eq:friedmann3}) reduces when we set $g_{1}=0$.

According to Eq.~(\ref{eq:thetaderne0}), the covariant 
divergence of the energy-momentum tensor~(\ref{eq:robwalk-T})
must be equal to the covariant derivative of the \emph{quadratic} Riemann tensor terms. Considering a classical content of the universe, namely dust and radiation, we can neglect torsion and assume metricity. 
For the zero component we then get
\begin{equation} \label{cc}
T\indices{^{\alpha}_{0\,;\,\alpha}}=-\dot{\rho}-3\frac{\dot{a}}{a}(\,\rho+p)=
g_{1}R\indices{_{\eta}^{\beta\lambda\alpha}_{;\alpha}}\,R\indices{^{\eta}_{\beta\lambda0}}.
\end{equation}

For the metric~(\ref{eq:robwalk}) only the zero component of the vector of the covariant
divergence of the quadratic Riemann tensor terms does not vanish, and with Eq.~ (\ref{eq:r-squared-div3}) we find
\begin{equation}\label{eq:eos2}
\onethird a\dot{\rho}+\dot{a}\left(\,\rho+p\right)=
2g_{1}\frac{\ddot{a}}{a^{3}}\left(-a^2\dddot{a}-a\dot{a}\ddot{a}+2\dot{a}^3+2\dot{a}K\right).
\end{equation}
Equation~(\ref{eq:eos2}) can be considerably simplified by inserting the trace equation~(\ref{eq:robwalk-trace}). With the definition of $M$ from Eq.~(\ref{eq:M-squared}) we finally get  (see appendix A) 
\begin{equation}\label{eq:deri-T}
\onethird \dot{\rho}+\dot{a}\left(\,\rho+p\right)=8\pi Gg_{1}\,\ddot{a}\left(\dot{p}-\onethird\dot{\rho}\right).
\end{equation}
This equation provides a simple relation between the density and the pressure and the scale parameter. As expected,  for $g_1 = 0$ we recover the covariant conservation of the stress energy tensor, as
\begin{equation}\label{eq:deri-T1}
\frac{d\rho}{dt}+3\frac{\dot{a}}{a}\left(\,\rho+p\right)=0
\end{equation}
corresponds to $T\indices{^{\alpha}_{0\,;\,\alpha}}=0$. On the other hand, in the case $g_1 \rightarrow \infty$, we obtain
\begin{equation}\label{eq:deri-T2}
\dot{p} = \frac{1}{3} \dot{\rho},
\end{equation}
which corresponds to the equation of state for radiation, Eq.~(\ref{eofr}). 

\medskip
After reviewing the asymptotic behavior of the CCGG-Friedman model, 
we devote the next section to analyzing the dynamics of the universe under the influence of matter represented by perfect fluids with  
various but constant equations of state.

\subsection{Rescaling the equations}
For solving the modified Friedman equations we express Eqs. (\ref{eq:friedmann}) in terms of the Hubble parameter,
\begin{equation}
H := \frac{\dot{a}}{a},
\end{equation}
and its derivatives. For simplicity and in alignment with observations we prefer to assume the space to be flat, i.e. 
$K=0$. The density and pressure equations are then rewritten as 
\begin{subequations}
\begin{equation}\label{density1}
8\pi G \rho^\prime = 3 H^2 + 24\pi G g_1\left(2H^2 \dot{H} + \dot{H}^2\right) 
\end{equation}
\begin{equation}\label{pressure1}
8\pi G p^\prime=-3 H^2 -2 \dot{H} + 8\pi G g_1\left(2 H^2 \dot{H}+\dot{H}^2\right)
\end{equation}
\end{subequations}
where the energy density and the pressure encompass dust, radiation and also include the cosmological constant interpreted as Dark Energy density and pressure, respectively:
\begin{subequations} \label{def:p+rho^prime}
\begin{equation}
\bar{\rho} = \rho + \frac{\Lambda}{8 \pi\, G}
\end{equation}
\begin{equation}
\bar{p} = p - \frac{\Lambda}{8 \pi\, G}
\end{equation}
\end{subequations}
This form reduces the modified Friedman equations from second order equations to first order equation. 
 \begin{figure*}[t!]
 \centering
\includegraphics[width=0.97\textwidth]{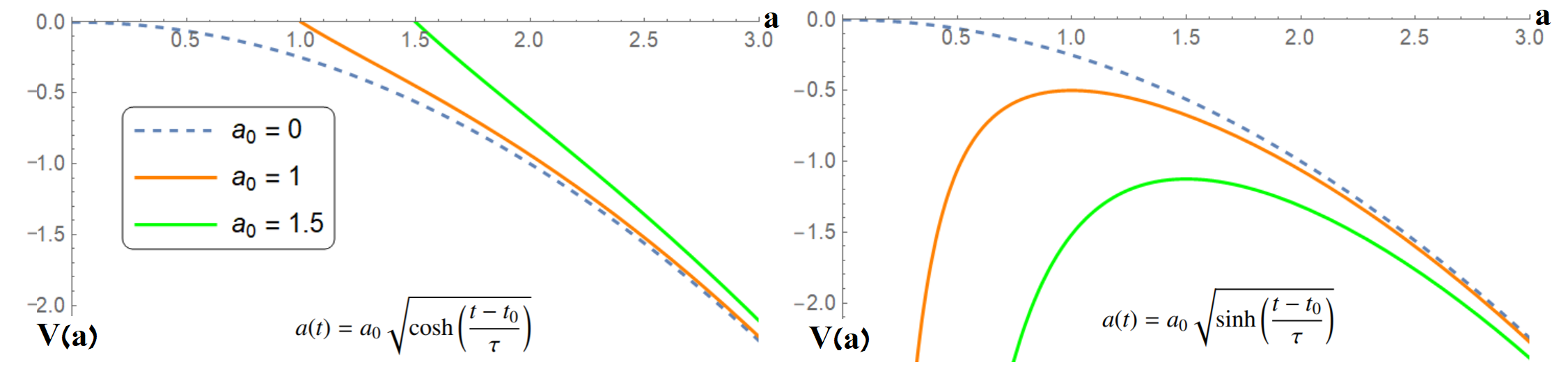}
\caption{The potential $V(a)$, Eq. \eqref{potential}, for two non-trivial solutions in the dark energy-dominant scenario, plotted vs. the scale parameter $a$. The solutions refer to Eqs. \eqref{ca} (left, $\cosh$) and \eqref{sa} (right, $\sinh$). The potential and the resulting scale dynamics depends on the integration constant   $a_0$. For $a_0 = 0$ both cases coincide.}
%
%
%
\label{fig1}
\end{figure*}
From the generalized Friedman equations, a typical time scale $\tau$ emerges related to the basic constant $g_1$:
\begin{equation}
\tau = \sqrt{8 \pi G g_1} = \sqrt{g_1}/M_p
\end{equation} where $M_p$ is the reduced Planck mass. This time scale is typical for the period of dominance of the quadratic term over the Einstein-Hilbert term in the full action. As we will see in the next section, for "short" times, $t/\tau \ll 1$, the quadratic term is  dominant and violates the conservation of the stress tensor. For long times, $t/\tau \gg 1$, the Einstein term becomes more dominant and asymptotically the universe behaves as the familiar cosmological model $\Lambda$CDM where the covariant conservation of the stress tensor is assumed. Notice, though, that the scale of $g_1$ can vary in the range  $0<g_1<10^{120}$ as explored in \cite{Vasak:2018gqn} which is equivalent to $0<\tau < 600 Gy$. Hence "short" and "long" are meant relative to this time scale!

The Friedman equations (\ref{eq:friedmann}) can be simplified by rescaling the involved quantities as 
\begin{align}\label{scale}
\tilde{t} := \frac{t}{\tau} 
\quad , \quad &\tilde{\rho} :=  (8 \pi G)^2 \,g_1\,\bar{\rho}  \quad  , \quad \tilde{H} := \tau\, H \\
&\tilde{p} :=  (8 \pi G)^2 \,g_1\,\bar{p}   \quad  , \quad \tilde{H}^\prime := \frac{d\tilde{H}}{d\tilde{t}} \nonumber
\end{align}
to get 
\begin{subequations}
\begin{equation}\label{density}
\tilde{\rho} = 3 \tilde{H}^2 + 3 \left(2\tilde{H}^2 \tilde{H}^\prime + \tilde{H}^{\prime 2}\right)
\end{equation}
\begin{equation}\label{pressure}
\tilde{p} =-3 \tilde{H}^2 -2 \tilde{H}^\prime + \left(\tilde{H}^{\prime 2} +2 \tilde{H}^2 \tilde{H}^\prime\right).
\end{equation}
\end{subequations}
The dependence on the constant $g_1$ has been absorbed by the new scale. For those equations, as we will see, for $\tilde{t} \gg 1$ the term $3\tilde{H}^2$ (which is familiar from the Einstein-Friedman model) would be more dominant for the density equation, and for $\tilde{t} \ll 1$ the term $3 \left(2\tilde{H}^2 \tilde{H}^\prime + \tilde{H}^{\prime 2}\right)$ (which comes from the quadratic Riemann term) would be more dominant.
 
In the following we suppress the tilde above the rescaled time coordinate and the other quantities and replace prime by dot for time derivatives, unless needed otherwise for clarity.
%
%

\section{Solutions with constant equations of state}
In this section we analyze three limiting cases of a \emph{radiation-dominated} (traceless stress tensor, $\omega = \frac{1}{3}$), \emph{dark energy dominated} ($\rho=-p$, $\omega = -1$),  and \emph{matter-dominated} ($p=0$, $\omega = 0$) universes.

A constant equation of state simplifies the rescaled generalized Friedman equations. By setting $p = \omega \rho$, and substituting Eq.~(\ref{pressure}) into Eq.~(\ref{density}), we obtain a unified equation which depends only on $\omega$ and on the Hubble parameter:
\begin{equation}\label{omega}
\tilde{H}' \left((3 \omega -1) \tilde{H}'+2\right)+\tilde{H}^2 \left(2 (3 \omega -1) \tilde{H}' +3 (\omega +1)\right) = 0
\end{equation}
Now it becomes obvious that for radiation ($\omega = \frac{1}{3}$) but also for dark energy ($\omega = -1$) a number of terms cancel out. 
\subsection{Radiation dominance}
For this case, there is no formal deviation from the trace equation, as the traces of both, the quadratic Riemann term (strain tensor) and the stress energy tensor vanish. Hence, the trace equation gives:
\begin{equation}
2\tilde{H}^2 + \tilde{H}'=0 \quad \Rightarrow \quad a \sim \tilde{t}^{\frac{1}{2}} \sim t^{\frac{1}{2}},
\end{equation}
recovering Eq. (\ref{omega}) with $\omega = \frac{1}{3}$
For this Hubble parameter, the new terms in Eqs.~(\ref{density}) and (\ref{pressure}) are identically zero:
\begin{equation}\label{vilation}
2\tilde{H}^2 \tilde{H}' + \tilde{H}^2 = 2H^2 H' + H^2 = 0.
\end{equation}
Hence both equations reduce to the original Friedman equation, and there are no deviations from the conventional radiation solution as all terms from the (traceless) quadratic Riemann tensor are identically zero. 
\subsection{Vacuum dominance}
Vacuum solution means a universe void of matter and radiation but with dark energy present. The equation of state is in this case $\omega = -1$ or $p=-\rho$.  Substituting this identity into the density and the pressure equations yields the differential equation
\begin{equation}\label{vaden}
\tilde{H}' \left(2 \tilde{H}'+4 \tilde{H}^2-1\right) = 0
\end{equation}
A trivial solution is $\dot{H}=0$ which is the standard inflation solution (see Ref.{\cite{Guth:1980zm}-\cite{Albrecht:1982wi}) for which, as in the radiation case (\ref{vilation}), the contribution from the quadratic Riemann tensor equals to zero. 
The energy density and the pressure are constant, hence we recover the standard Einstein Ansatz with cosmological constant. 
Notice, though, that by rescaling, the cosmological constant
\begin{equation}
\Lambda \rightarrow \Lambda \, (8 \pi G)^2\,g_1
\end{equation}
grows with growing $g_1$, the clock ticks faster via $\tilde{t} \sim t/\sqrt[]{g_1}$.
 \begin{figure*}[t]
 	\centering
\includegraphics[width=1\textwidth]{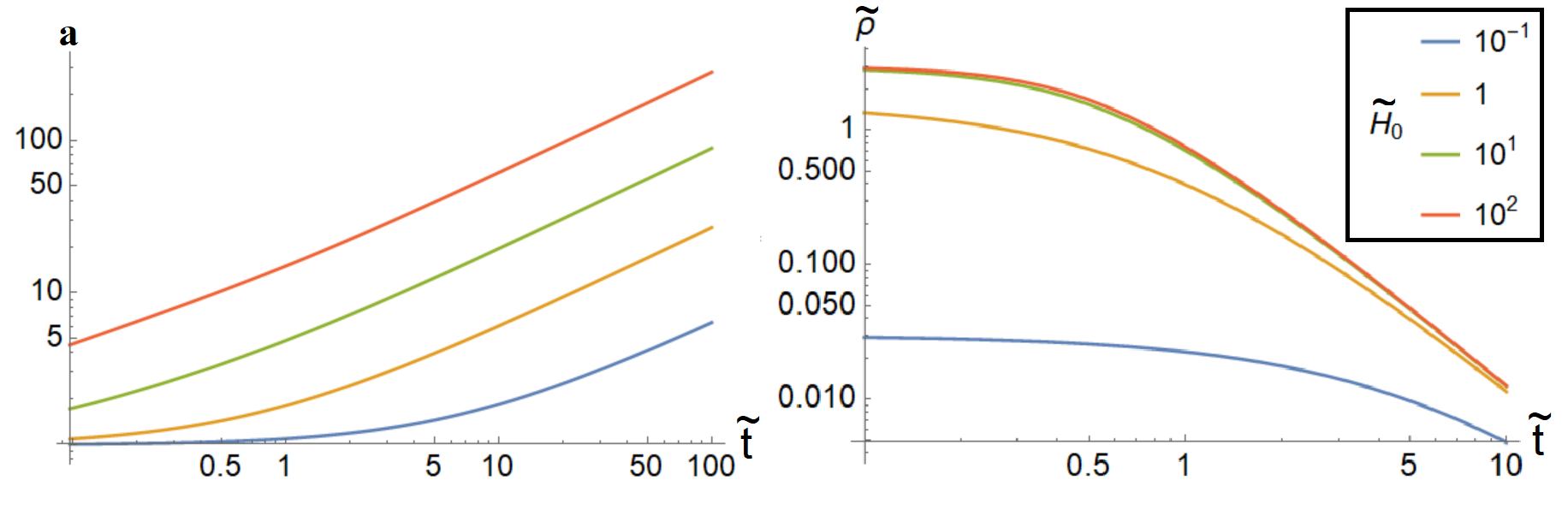}
\caption{The scale parameter and the energy density for a matter-dominated universe as a function of the rescaled time coordinate. }
%
%
 	\label{fig2}
 \end{figure*} 
The second solution, in the vacuum dominant case, results if the expression in parentheses vanishes:
\begin{equation}
2 \tilde{H}'+4 \tilde{H}^2-1=0.
\end{equation} 
The solution for the scale factor is readily determined with two different solutions:
\begin{equation}\label{ca}
a(t)= a_{0} \sqrt{\cosh\left(\frac{t-t_0}{\tau}\right)}
\end{equation}
for $|\tilde{H}|  < 1/2$
or:
\begin{equation}\label{sa}
a(t)= a_{0} \sqrt{\sinh\left(\frac{t-t_0}{\tau}\right)}
\end{equation}
for $|\tilde{H}| > 1/2$.
Here $a_{0}$ and $t_0$ are integration constants. For $t=t_0$ the scale parameter in (\ref{ca}) denotes the minimal value $a_{min} = a_{0}$. This solution which is symmetric with respect to $t-t_0$ describes a bouncing universe \cite{Steinhardt:2001vw}-\cite{Steinhardt:2001st}.  Starting from minus infinite time it  decelerates to a full stop, at time $t=t_0$ and finite scale factor $a_{0}$, to rebound thereafter into an exponential inflation phase.
For the second solution (\ref{sa}), the universe starts with $a=0$ which describes a Big Bang with asymptotic  exponential inflation. 

The energy density and pressure in this case can be investigated by substituting the scale parameter into Eqs.~(\ref{density}) and~(\ref{pressure}). Surprisingly, the energy density and pressure remain constant. Therefore from the quadratic Riemann term a new inflationary solution emerges, and the special vacuum solution still leads to a constant density with the equation of state of $\omega = -1$.    
%
%
In order to illuminate the physical behavior of the vacuum-dominant case  we have restored the potential of the scale factor dynamics from Eqs. \eqref{ca} and \eqref{sa} by demanding that the "kinetic energy" $\dot{a}^2$  and the potential energy $V(a)$ would add up to zero. By differentiation we obtain:
\begin{equation}\label{potential}
\dot{a}^2 + V(a) = 0, \qquad V(a) =  -\frac{a^2}{4}\pm\frac{a_{0}^4}{4 a^2}
\end{equation}
The sign "$+$" refers to the first, symmetric solution, \eqref{ca}, and the "$-$" sign refers to the second, antisymmetric solution, (\ref{sa}). 
The graph for those potentials is shown in Fig.~(\ref{fig1}). The bouncing solution with a finite minimum scale $a_{0}$ for the symmetric, $cosh$, case is displayed in the $\textbf{left}$ figure. In contrast, the antisymmetric, $sinh$, solution ($\textbf{right}$ figure) has a singularity in the origin. The asymptotic form of the  potential is $V(a) \sim -a^2$, in both cases, such that the late epoch solution will be $\dot{a} / a = H = \mathrm{const}$ which is the standard inflationary solution and applies here for both, the vacuum  and radiation solutions. For a matter-dominated universe the picture is different, though.

\subsection{Matter dominance}
\begin{figure}[b]
\centering\includegraphics[width=0.5\textwidth]{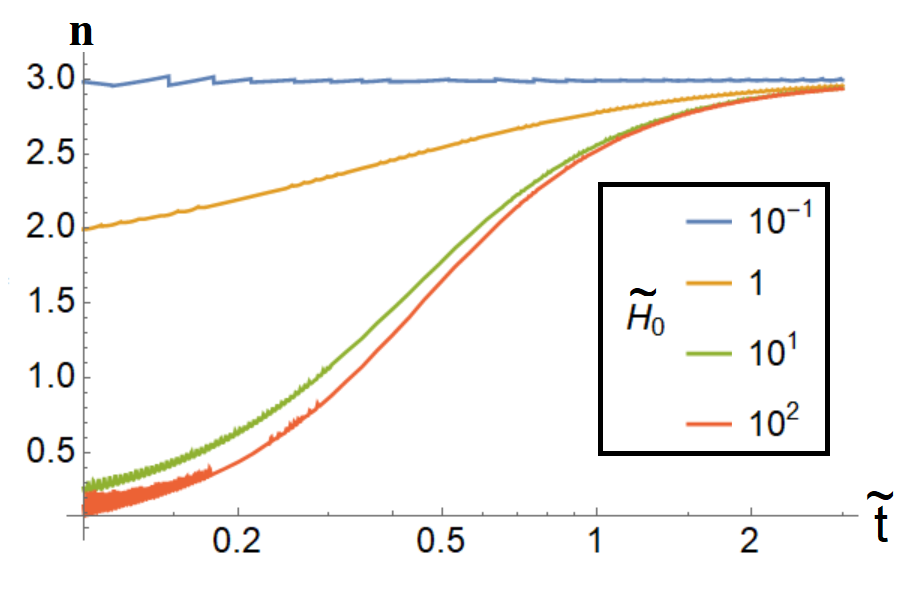}
\caption{The Power of the scale factor which gives the density vs. the rescaled time.}
\label{fig3}
\end{figure} 
In the matter-dominated case, the second Friedman equation, (\ref{pressure}), yields the differential equation for the Hubble parameter:
\begin{equation}\label{DF2}
3 \tilde{H}^2 + 2 \tilde{H}' = 2\tilde{H}^2\tilde{H}'+\tilde{H}'^2,
\end{equation}
where  the time refers again to the rescaled coordinate $\tilde{t}$. Numerical solutions for different values of  the initial condition $H_0 \equiv H(t=0)$ are presented in Fig. (\ref{fig2}). The initial values $H_0$ are selected across orders of magnitude to provide a feeling for the general behavior of the solution. The units of the Hubble parameters are $1/\tilde{\tau}$.

To see the physics in between different values of $g_1$, we define the quantity:
%
\begin{equation}\label{power}
\rho \sim \frac{1}{a^n}\quad\Leftrightarrow\quad n=-\frac{1}{H}\frac{d}{dt}\ln{\rho}
\end{equation}
which gives the power dependence of the density versus the power of the scale factor for a power law universes, which for matter dominant is $n = 3$ and $n = 4$ is for radiation.

By this redefinition, the numerical solutions in Fig.~(\ref{fig3}) show the difference between physical evolution for different values of $H_0$. With larger values of $H_0$ the  deviation of the density scaling from $n=3$, which is the standard dust solution, is stronger in the short term. However, all solutions go asymptoticly to the power $n=3$ in the long term.   This is consistent with the evolution of the  energy momentum tensor. Fig.~(\ref{fig4})  shows the  covariant derivative of the stress energy tensor,  numerically calculated from Eq.~(\ref{cc}), versus the re-scaled  time $\tilde{t}$.  We observe that the covariant conservation law for the stress tensor holds  asymptotically  for $t \gg \tau$.  At those times the Riemann square term from the action fades away leaving just the conventional linear Einstein term.  However, with increasing value of the time scale $\tau$  the time where  the covariant derivative  approaches zero is delayed. 
For illustrating the energy transition between space-time and matter 
we analyze the quantity $\Omega_m^{(0)} = \frac{\tilde{\rho}_m(a)}{\tilde{\rho}_m(0)}  a^3 $. In standard cosmology this quantity is a constant, but in our case it will be constant only asymptotically, for $\tilde{t} \gg 1$, as  we can see from Fig.~(\ref{fig6}). The normalized energy density $\Omega_m$ converges to  $1$  in the "late epoch". In earlier times energy density is gradually transferred from  space-time (especially the quadratic Riemann term) to matter up to the value we encounter in the late epoch.
%
%
 \begin{figure}[t]
\centering\includegraphics[width=0.5\textwidth]{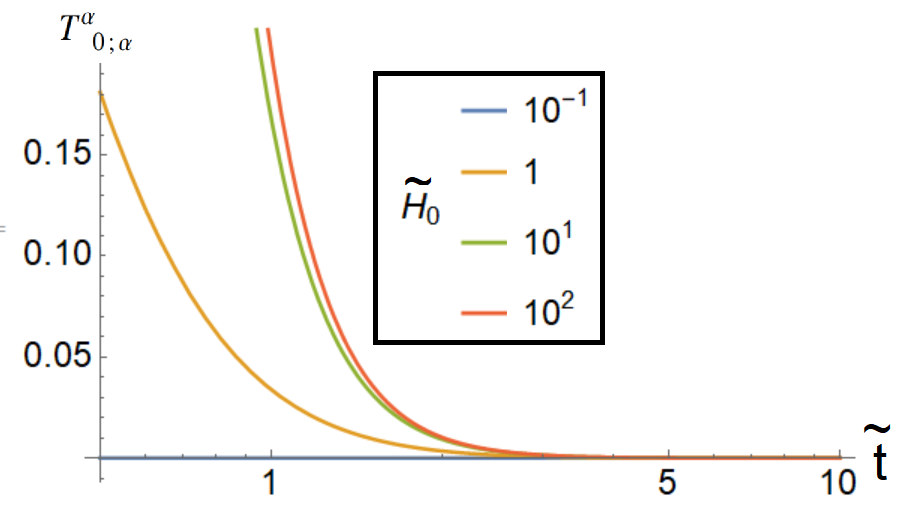}
\caption{Covariant derivative of the stress energy tensor vs. the rescaled time.}
\label{fig4}
\end{figure} 
\section{Conclusions\label{sec:conclusions}}
In this paper we have investigated the impact of the covariant canonical gauge theory of gravity on simple cosmological scenarios. Because this formulation is based on the first order formalism, the energy-momentum is not covariantly conserved -- as well known also for other theories beyond Lovelock. We interpret this as energy transfer from space-time to matter. For a cosmological solution in a homogeneous and isotropic universe, the Friedman equations are modified. For $g_1 = 0$ we recover the original Friedman equations, and for $g_1 \rightarrow \infty$ the modification becomes dominant. The equation of state for this limit is similar to the equation of state of radiation. The coupling constant $g_1$ driving the strength of the quadratic term gives rise to a time scale $\tau=\sqrt{8 \pi G g_1}$. For $t \ll \tau$ the quadratic Riemann term becomes dominant, whereas for $t \gg \tau$ the Einstein term is dominant. Therefore, a deviation from the standard cosmology emerges only in the very early universe. Of course "early" refers to the  time scale driven by the coupling constant $g_1$.

We have analyzed various scenarios of the Friedman universe filled with perfect fluids with constant equations of state. For pure radiation we find  no deviation from the standard cosmological solution, as the new quadratic term is traceless alike the radiation stress tensor. For a dark energy solution the original inflationary solution is predicted, but in addition we get a bouncing  and a Big Bang solution, both with asymptotic inflation. 
\begin{figure}[b!]
\centering\includegraphics[width=0.5\textwidth]{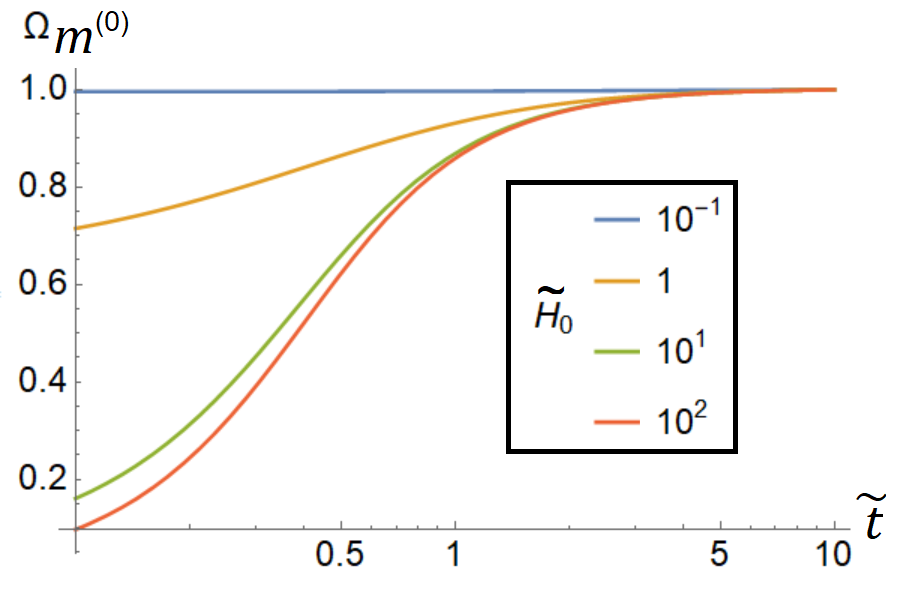}
\caption{The $\Omega_{m}^{(0)}$ vs. the normalized time.}
\label{fig6}
\end{figure}
The dust solution is the most interesting one. The deviation from the standard Friedman equation arises only for $t \ll \tau$ where the covariant conservation of the stress energy tensor is not zero.  In fact, because the Einstein tensor in the stress tensor is covariantly conserved, the quadratic Riemann and the matter density terms must be conserved together. This implies the interpretation that energy-momentum is transferred from  space-time to matter. Asymptotically the solution settles at the standard dust solution. 

In summary, the effect of the quadratic Riemann term in the CCGG equation leads to a modified dynamics of space-time and matter. Derived via the rigorous mathematical framework of canonical transformations from first principles, we encounter "Dark Energy" like effects that have its roots in the dynamics of the geometry of the universe. A coherent geometrical theory of Dark Energy 
from the CCGG formulation is subject of an ongoing study. Another possibility to maintain the energy momentum tensor conservation from different aspects of CCGG, without torsion described in \cite{Benisty:2018efx}\cite{Benisty:2018ufz}
 In the future we will study cosmological solutions which include the possibility of torsion, and in this way we could possibly maintain the energy momentum conservation. 
\section{Appendix - Covariant divergence of the quadratic term}
We first set up directly the covariant derivative of the quadratic Riemann tensor expression from Eq.~(\ref{eq:r-squared})
\begin{align}
Q\indices{^{\alpha}_{\xi;\alpha}}&:=
\left(R\indices{^{\,\eta\beta\lambda\alpha}}\,R\indices{_{\eta\beta\lambda\xi}}-
\quarter\delta_{\xi}^{\alpha}R\indices{^{\,\eta\beta\lambda\tau}}\,R\indices{_{\eta\beta\lambda\tau}}\right)_{;\alpha}\nonumber\\
&=R\indices{^{\,\eta\beta\lambda\alpha}_{;\alpha}}\,R\indices{_{\eta\beta\lambda\xi}}+
R\indices{^{\,\eta\beta\lambda\alpha}}\,R\indices{_{\eta\beta\lambda\xi;\alpha}}-
\onehalf R\indices{^{\,\eta\beta\lambda\tau}}\,R\indices{_{\eta\beta\lambda\tau;\xi}}\nonumber\\
&=R\indices{^{\,\eta\beta\lambda\alpha}}\left(R\indices{_{\eta\beta\lambda\xi;\alpha}}-
\onehalf R\indices{_{\eta\beta\lambda\alpha;\xi}}\right)+
R\indices{^{\,\eta\beta\lambda\alpha}_{;\alpha}}\,R\indices{_{\eta\beta\lambda\xi}}.
\label{eq:r-squared-div}
\end{align}
We now make use of the Bianchi identity for the covariant derivative of the Riemann tensor in spaces without torsion
\begin{equation*}
R\indices{_{\eta\beta\lambda\alpha;\xi}}+R\indices{_{\eta\beta\alpha\xi;\lambda}}+R\indices{_{\eta\beta\xi\lambda;\alpha}}=0,
\end{equation*}
which we insert into Eq.~(\ref{eq:r-squared-div})
\begin{equation}
Q\indices{^{\alpha}_{\xi;\alpha}}
=R\indices{^{\,\eta\beta\lambda\alpha}}\left(-R\indices{_{\eta\beta\xi\lambda;\alpha}}+
\onehalf R\indices{_{\eta\beta\alpha\xi;\lambda}}+\onehalf R\indices{_{\eta\beta\xi\lambda;\alpha}}\right)\nonumber
+R\indices{^{\,\eta\beta\lambda\alpha}_{;\alpha}}\,R\indices{_{\eta\beta\lambda\xi}}.
\end{equation}
and equivalently rewritten as
\begin{equation*}
Q\indices{^{\alpha}_{\xi;\alpha}}=-\onehalf\cancel{R\indices{^{\,\eta\beta\lambda\alpha}}\left(R\indices{_{\eta\beta\xi\lambda;\alpha}}+R\indices{_{\eta\beta\xi\alpha;\lambda}}\right)}+R\indices{^{\,\eta\beta\lambda\alpha}_{;\alpha}}\,R\indices{_{\eta\beta\lambda\xi}}.
\end{equation*}
The first term on the right-hand side vanishes as the sum of the derivatives of the Riemann tensor
is symmetric in $\alpha$ and $\lambda$ while $R\indices{^{\,\eta\beta\lambda\alpha}}$ is skew-symmetric in these indices.
The final result is now
\begin{equation}\label{eq:r-squared-div3}
R\indices{_{\eta}^{\beta\lambda\alpha}_{;\alpha}}\,R\indices{^{\eta}_{\beta\lambda 0}}=
6\frac{\ddot{a}}{a^{4}}\left(a^2\dddot{a}+a\dot{a}\ddot{a}-2\dot{a}^3-2\dot{a}K\right).
\end{equation}
Hence
\begin{equation}
\begin{split}
\onethird a^{4}\dot{\rho}+a^{3}\dot{a}\left(\,\rho+p\right) 
=2g_{1}a\ddot{a}\ddfrac{}{t}(\dot{a}^2-2Ma^{2}+K)\\+4g_{1}\dot{a}\ddot{a}(2Ma^{2}-a\ddot{a})=-4g_{1}a^{3}\ddot{a}\ddfrac{}{t}M
\end{split}
\end{equation}

\end{document}